\documentclass[preprint,aps,tightenlines,superscriptaddress]{revtex4}
\usepackage{graphicx}
\usepackage{bm}
\usepackage{epsfig}
\newif\ifpdf
\ifx\pdfoutput\undefined
\pdffalse 
\else
\pdfoutput=1 
\pdftrue
\fi

\newcommand{\bea}{\begin{eqnarray}}
\newcommand{\eea}{\end{eqnarray}}
\newcommand{\beq}{\begin{equation}}
\newcommand{\eeq}{\end{equation}}
\newcommand{\bay}{\begin{array}}
\newcommand{\eay}{\end{array}}
\newcommand{\Dslash}{\mbox{$\not{\hspace{-1.03mm}D}$}}        

\begin{document}
\ifpdf
\DeclareGraphicsExtensions{.pdf, .jpg}
\else
\DeclareGraphicsExtensions{.eps, .jpg}
\fi
\vspace{1.5cm}
\preprint{ \vbox{\hbox{UCSD/PTH 02-02} \hbox{hep-ph/0201298}   }}
\vspace{2.0cm}

\title{Symmetry-breaking corrections to heavy meson form-factor relations} 
\author{Benjam\'\i{}n Grinstein and Dan Pirjol}
\affiliation{Department of Physics, UCSD, 9500 Gilman Drive, La Jolla, CA 92093}

\date{\today\\ \vspace{2cm} }

\begin{abstract}
\vspace{1.0cm}
\setlength\baselineskip{18pt}

In the heavy quark limit, the form factors for semileptonic and rare radiative
$B$ decays into light mesons are related by heavy quark spin symmetry. 
Here we compute the leading corrections of order $\Lambda/m_Q$ to these symmetry 
relations, showing also how to include hard gluon effects systematically
to any order in $\alpha_s(m_Q)$. 
The subleading correction to the form factor relation for $B\to \pi$ can be
computed exactly in the soft pion limit. For decays into vector mesons, 
the matrix elements of two local dimension-4 operators are needed,
one of which vanishes in the constituent quark model.
A few applications are briefly discussed.

\end{abstract}

\maketitle


A precise determination of the Cabibbo-Kobayashi-Maskawa matrix 
elements using exclusive heavy meson decays depends in a crucial 
way on our theoretical control over strong interaction effects 
in these processes. The heavy quark effective theory (HQET) \cite{MaWi} 
provides a systematic framework within which the relevant matrix elements
can be studied in an expansion in powers of $1/m_Q$, the inverse heavy
quark mass. We focus in this Letter on form factors for heavy-light
transitions, which are relevant for semileptonic and rare decays such as
$\bar B\to \rho\ell\bar\nu_\ell$ and $\bar B\to K^*\ell^+\ell^-$.

Heavy quark spin symmetry has been used to relate heavy-to-light form
factors corresponding to different currents \cite{IW,BuDo}. These relations 
hold true 
in the kinematical region where the energy of the final light hadron $E$ in the
rest frame of the $B$ meson is not too large. At large recoil 
$E\gg \Lambda_{QCD}$, a new 
symmetry comes into play, further simplifying the structure of the form 
factors \cite{LEET}. At the same time, ${\cal O}(\alpha_s(m_Q))$ perturbative 
corrections due to the exchange of hard gluons become important. 
They can be computed in a systematic way in perturbation theory, and the
corresponding ${\cal O}(\alpha_s)$ corrections have been
evaluated in \cite{BeFe,BFPS}.

In both these kinematical regions there are corrections of order $\Lambda/m_Q$ 
to the symmetry relations for the form factors, arising from subleading operators
in the HQET. The complete expansion of the individual form factors at 
order $\Lambda/m_Q$ involves in general a large number of both local and
nonlocal contributions and appears to preserve little predictive power.
Such analyses have been in given in \cite{BLNN} for $\bar B\to P$ and
in \cite{Hu} for the $\bar B\to V$ form factors, where $P$ and $V$ are
pseudoscalar and vector light mesons, respectively. We show in this 
Letter that, with certain modifications required to include hard gluon 
corrections,
the symmetry relations can be simply extended to subleading order. 
The subleading corrections have a simple form and can be expressed only
in terms of lowest order form factors and the matrix elements of two 
dimension-4 local operators.\\

The hadronic matrix elements relevant for $\bar B$ decays into a light meson
are parametrized in terms of form factors. There are three form factors for
decays into a pseudoscalar, relevant for semileptonic decays (e.g.
$B\to \pi e\nu$)
\bea\label{P1}
\langle P(p')|\bar q\gamma_\mu b|\bar B(p)\rangle =
f_+ (p_\mu + p'_\mu) + f_- (p_\mu - p'_\mu)
\eea
and rare decays (e.g. $B\to K e^+ e^-$)
\bea\label{P2}
\langle P(p')|\bar q\sigma_{\mu\nu} b|\bar B(p)\rangle =
is[(p_\mu + p'_\mu)(p_\nu - p'_\nu)  - (p_\nu + p'_\nu)(p_\mu - p'_\mu)]
\eea

The $\bar B$ decays into a light vector meson are parametrized by a total 
of seven form factors, four of which appear in  semileptonic decay matrix 
elements
\bea\label{VA}
\langle V(p',\epsilon)|\bar q\gamma_\mu b|\bar B(p)\rangle &=&
ig(q^2) \varepsilon_{\mu\nu\lambda\sigma} \epsilon^*_\nu
(p+p')_\lambda (p-p')_\sigma\\
\label{4}
\langle V(p',\epsilon)|\bar q\gamma_\mu\gamma_5 b| \bar B(p)\rangle &=&
f(q^2) \epsilon^*_\mu \\
&+& \, a_+(q^2)(\epsilon^*\cdot p)(p+p')_\mu +
a_-(q^2)(\epsilon^*\cdot p)(p-p')_\mu\,.\nonumber
\eea
and three form factors in rare radiative decays
\bea\label{T}
\langle V(p',\epsilon)|\bar q\sigma_{\mu\nu} b|\bar B(p)\rangle &=&
g_+(q^2) \varepsilon_{\mu\nu\lambda\sigma} \epsilon^*_\lambda
(p+p')_\sigma + 
g_-(q^2) \varepsilon_{\mu\nu\lambda\sigma} 
\epsilon^*_\lambda (p-p')_\sigma\\
 & &+
h(q^2) \varepsilon_{\mu\nu\lambda\sigma} (p+p')_\lambda
(p-p')_\sigma(\epsilon^*\cdot p)\nonumber\,.
\eea
While the vector and axial form factors are renormalization scale invariant,
the form factors of the tensor current have a nontrivial scale dependence.
Throughout in the following it will be understood implicitly that the scale 
used to define these form factors is $\mu = m_b$.

Counting powers of $m_b$ coming from kinematical factors and the usual relativistic 
normalization of the $|\bar B(p)\rangle$ states gives the heavy quark mass
scaling properties of these form factors \cite{IW}
\beq\label{4.9}
f_+(y) + f_-(y) \propto m_b^{-1/2}\,,\qquad 
f_+(y) - f_-(y) \propto m_b^{1/2}\,,\qquad s(y) \propto m_b^{-1/2}
\eeq
and
\beq\label{5}
f(y) \propto m_b^{1/2}\,,\quad
g(y) \propto m_b^{-1/2}\,,\quad a_+(y) - a_-(y) \propto m_b^{-1/2}\,,\quad
a_+(y) + a_-(y) \propto m_b^{-3/2}
\eeq
\beq\label{5.1}
g_+(y) - g_-(y) \propto m_b^{1/2}\,,\quad
g_+(y) + g_-(y) \propto m_b^{-1/2}\,,\quad h \propto m_b^{-3/2}\,.
\eeq
These scaling relations hold at a fixed value of the light meson energy
in the rest frame of $B$, $E_V = m_V y$. Therefore for a consistent $m_b$ power
counting, we will use $y=p\cdot p'/(m_B m_V)$ as argument for the form factors,
instead of $q^2$. These parameters are related as $q^2 = m_B^2 + m_V^2 - 
2m_B m_V y$.
The range of $y$ probed in 
semileptonic $D$ decays is $1 \leq y \leq y_D^{\rm max} = 
\frac{m_D^2 + m_V^2}{2m_D m_V} \simeq 1.3$, whereas in $B$ decays the
corresponding range is $1 \leq y \leq 3.0$.

For values of $y=v\cdot v'$ not too far away from the zero recoil point 
$y=1$, heavy quark spin symmetry can be used to relate some of the form 
factors in (\ref{P1})-(\ref{T}) \cite{IW,BuDo}. 
There is one relation among the $B\to P$ form factors
\bea\label{IV}
\mbox{(P-1)}:\qquad f_+(y) - f_-(y) - 2m_B s(y) = {\cal O}(m_b^{-1/2})
\eea
and three relations among the $B\to V$ form factors
\bea\label{I}
\mbox{(V-1)}:\qquad & & g_+(y) - g_-(y) = -2m_B g(y) + {\cal O}(m_b^{-1/2})\\
\label{II}
\mbox{(V-2)}:\qquad & & g_+(y) + g_-(y) = 2m_V y g(y) + 
\frac{1}{m_B}f(y) + {\cal O}(m_b^{-3/2})\\
\label{III}
\mbox{(V-3)}:\qquad & & a_+(y) - a_-(y) = 2 g(y) - 2m_B h  + {\cal O}(m_b^{-3/2})\,.
\eea
In the following we will review the derivation of these relations
using a method which will allow us to include hard gluon corrections
to all orders in $\alpha_s(m_b)$, and to compute the
leading $O(\Lambda/m_b)$ corrections to them.\\

We illustrate the principle of the method 
on the example of Eq.~(\ref{IV}), and start by considering the two currents
$J_\mu = \bar q\gamma_\mu b$ and $J'_\mu = i\bar q\sigma_{\mu\nu} v^\nu b$.
Both of them are matched in HQET onto operators of the general form
\bea\label{Jeff}
J_\mu^{(\prime )} =
\bar q\Gamma^{(\prime )}_\mu b = 
c_0^{(\prime )}(\mu) \bar q\gamma_\mu h_v + 
c_1^{(\prime )}(\mu) \bar qv_\mu h_v
+ \frac{1}{2m_b}\sum_{i=1}^6 b^{(\prime )}_i(\mu) {\cal J}_\mu^i\,,
\eea
with ${\cal J}^i_\mu$ the most general set of dimension-4 operators with the
same transformation properties as the corresponding QCD current.
Restricting ourselves to operators which do not 
vanish by the equations of motion, a complete basis can be chosen 
as
\bea\label{Ji}
{\cal J}^1_{\mu} &=& \bar q\gamma_\mu i\Dslash h_v\,,\qquad\quad
{\cal J}^2_{\mu} = \bar qv_\mu i\Dslash h_v\,,\qquad\,\,
{\cal J}^3_{\mu} = \bar q iD_\mu h_v\,,\\
{\cal J}^4_{\mu} &=& v\cdot i\partial (\bar q\gamma_\mu h_v)\,,\quad
{\cal J}^5_{\mu} = v\cdot i\partial (\bar q v_\mu h_v)\,,\quad
{\cal J}^6_{\mu} = i\partial_\mu (\bar q h_v)\,.\nonumber
\eea

The Wilson coefficients of the dimension-3 operators 
$c_{0,1}^{(\prime )}(m_b)$ 
at the matching scale  $c_i(m_b)$ are known to NNL order \cite{BrGr}.
For $J_\mu$ they are given by
\bea\label{cmatch}
& &c_0(m_b) =1 - \frac{\alpha_s C_F}{\pi} - (11.40-0.79 n_\ell + 0.09)
\left(\frac{\alpha_s}{\pi}\right)^2\\
& & c_1(m_b) = \frac{\alpha_s C_F}{2\pi} + (7.30-0.35 n_\ell+0.02)
\left(\frac{\alpha_s}{\pi}\right)^2 \,,
\eea
and for $J'_\mu$ by
\bea\label{c'match}
& &c'_0(m_b) =- c'_1(m_b) = -(1 - \frac{\alpha_s C_F}{\pi} - (16.09
- 1.13 n_\ell + 0.13)\left(\frac{\alpha_s}{\pi}\right)^2)\,.
\eea
The corresponding Wilson coefficients of the dimension-4 operators
are given at tree level by
\bea\label{bmatch}
& &b_1(m_b) = 1\,,\quad b_{2-6}(m_b) = 0\\
& &b'_1(m_b) = b'_2(m_b) = 1\,,\quad  b'_{3-6}(m_b) = 0\,.\nonumber
\eea
In addition to the local corrections coming from the matrix elements
of the ${\cal J}^i_\mu$ operators, there are also nonlocal subleading
$1/m_b$ operators which appear as time-ordered products of the 
leading dimension-3 currents with the $1/m_b$ terms in the HQET
Lagrangian
\bea
{\cal L}_{\rm kin} = \bar h_v (iD)^2 h_v\,,\qquad 
{\cal L}_{\rm mag} = \frac12 \bar h_v g\sigma_{\mu\nu} F^{\mu\nu} h_v\,.
\eea

Reparametrization invariance \cite{LuMa} gives the relation 
$c_0(\mu) = b_1(\mu)$ among the Wilson 
coefficients of the current $J_\mu$, which should be satisfied to all orders 
in $\alpha_s$.
No such constraints exist among the Wilson coefficients of $J'_\mu$, 
for which reparametrization invariance has been broken 
explicitly by defining it in terms of $v_\mu$ (the 
corresponding constraints on the Wilson coefficients of $\bar c\sigma_{\mu\nu}b$
are not sufficiently predictive to relate $b'_i$ and $c'_i$). However, the 
condition $v^\mu J'_\mu = 0$ gives the 
following constraints for $c'_i(\mu)$ and $b'_i(\mu)$ separately
\bea\label{Jconst}
c'_0(\mu) = -c'_1(\mu)\,,\qquad b'_1(\mu) = b'_2(\mu)\,,\quad 
b'_3(\mu) + b'_4(\mu) + b'_6(\mu) = 0\,.
\eea

The symmetry relation (\ref{IV}) follows from taking an appropriate linear combination
of matrix elements of $J_\mu$ and $J'_\mu$ between hadronic states 
$\langle \pi(p') |J_\mu + \kappa_1 J'_\mu|\bar B(p)\rangle$, 
with $\kappa_1$ to be chosen as explained below. 
In the full theory this matrix element can be expressed in terms of the form
factors (\ref{P1}), (\ref{P2})
\bea\label{fullIV}
& &\langle \pi(p')|J_\mu + \kappa_1 J'_\mu |
\bar B(p)\rangle = \\
& &\qquad\qquad
[ f_+ - \kappa_1 (m_B - v\cdot p') s] (p+p')_\mu +
[ f_- + \kappa_1 (m_B - v\cdot p') s] (p-p')_\mu \nonumber
\eea
The same matrix element is given in the effective theory by 
\bea\label{effIV}
& &\langle \pi(p')|J_\mu + \kappa_1 J'_\mu |
\bar B(p)\rangle = (c_0 + \kappa_1 c'_0) \langle \bar q\gamma_\mu h_v \rangle + 
(c_1 + \kappa_1 c'_1) \langle \bar q v_\mu h_v \rangle\\
& &\quad + \frac{1}{2m_b}\left\{
\sum_{a={\rm kin, mag}}
c_{\rm a}(\mu) (c_0(\mu)+\kappa_1 c'_0(\mu)) 
\langle T\{\bar q\gamma_\mu h_v,\, i{\cal L}_a\}\rangle\right.\nonumber\\
& &\quad\left. +
c_{\rm a}(\mu) (c_1(\mu)+\kappa_1 c'_1(\mu)) 
\langle T\{\bar qv_\mu h_v,\, i{\cal L}_a\}\rangle +
 \sum_{i=1}^6(b_i + \kappa_1 b'_i) \langle {\cal J}^i_\mu\rangle
\right\}\nonumber
\eea

At tree level, the combination of Wilson coefficients $c_0 + \kappa_1 c'_0$
multiplying the matrix element of $\bar q\gamma_\mu h_v$
vanishes (with $\kappa_1 = 1$). Since the contribution of $\bar q v_\mu h_v$ is
proportional to $p_\mu$, it follows that the combination of form factors
multiplying $p'_\mu$ in (\ref{fullIV}) is suppressed by $\Lambda/m_b$. 
This was the original observation of \cite{IW} leading to the symmetry 
relation (P-1) (\ref{IV}).
From (\ref{cmatch}) and (\ref{c'match}) one can see that $c_0 + c'_0=0$ is accidentally
satisfied also at one-loop order, and a correction appears first at two-loop order.
We will take in the following
\beq\label{kappa1}
\kappa_1 = -\frac{c_0(m_b)}{c'_0(m_b)} = 1 + (4.69 - 0.34 n_\ell + 0.04)\left(
\frac{\alpha_s(m_b)}{\pi}\right)^2
\eeq
such that this cancellation persists to any order in $\alpha_s(m_Q)$.
Furthermore, this ensures also the cancellation of the 
nonlocal power-suppressed terms $\langle {\cal L}_{\rm a}, \bar q\gamma_\mu h_v
\rangle$. 
The remaining dimension-4 local terms can be easily expressed in terms of leading
order form factors plus the additional form factors $\delta_\pm(y)$ defined as
\bea\label{ddef}
\langle \pi(p',\epsilon) |\bar q iD_\mu h_v |\bar B(v)\rangle =
\delta_+(y) (p_\mu + p'_\mu) + \delta_-(y) (p_\mu - p'_\mu)\,.
\eea
The equation of motion for the $h_v$ field gives one constraint among these
form factors $(m_B + v\cdot p') \delta_+(y) + (m_B - v\cdot p') \delta_-(y) =0$,
such that only one of them is independent. Inserting this into (\ref{effIV}) and comparing
the coefficient of $p'_\mu$ on both sides, one finds the improved version of the 
symmetry relation (\ref{IV}) 
\bea\label{IVcorr}
\mbox{(P-1$'$)}:\qquad
f_-(y)  + \kappa_1 m_B s(y) = -\frac{1}{m_B}(\delta_+ - \delta_-) - 
(\bar\Lambda - \kappa_1 v\cdot p') s(y) + {\cal O}(m_b^{-3/2})
\eea
We used on the right-hand side of (\ref{effIV}) the tree level values of the Wilson 
coefficients  $b^{(\prime )}_i(m_b)$
(\ref{cmatch}), (\ref{c'match}), which is sufficient for the accuracy intended. 
The corrections to (\ref{IVcorr}) are suppressed by $\frac{\Lambda^2}{m_Q^2}$ 
relative to the leading terms in (\ref{IV}), and come 
from matrix elements of dimension-5 operators which were neglected 
in (\ref{effIV}).\\

Similar improved relations can be proved for the $\bar B\to V$ form factors. 
The first
symmetry relation (\ref{I}) follows again from taking the matrix element
\bea\label{sumQCD}
\langle \rho(p',\epsilon) | J_\mu  + 
\kappa_1 J'_\mu |\bar B(v)\rangle =
-\left( 2g + \kappa_1 \frac{1}{m_B} (g_+ - g_-)\right)i \varepsilon_{\mu\nu\lambda\sigma}
\epsilon_\nu^* p_\lambda p'_\sigma\,.
\eea
The same quantity can be expressed in the effective theory with the help of (\ref{Jeff})
as
\bea\label{sumeff}
\langle \rho | J_\mu + \kappa_1 J'_\mu |\bar B\rangle &=&
(c_0(\mu) + \kappa_1 c'_0(\mu)) \langle \bar q\gamma_\mu h_v\rangle\\
& &\hspace{-4cm} +\frac{1}{2m_b}\left\{\sum_{a={\rm kin, mag}}
c_{\rm a}(\mu) (c_0(\mu)+\kappa_1 c'_0(\mu)) 
\langle T\{\bar q\gamma_\mu h_v,\, i{\cal L}_a\}\rangle +
 \sum_{i=1}^6(b_i + \kappa_1 b'_i) \langle {\cal J}^i_\mu\rangle  
\right\}\nonumber
\eea
The matrix element of $\bar q v_\mu h_v$ vanishes for this case by Lorentz
invariance. The nonlocal $1/m_b$ contributions cancel as before, leaving 
only local dimension-4 operators. They can be easily expressed in terms of 
leading order form factors plus one additional form factor ${\cal D}(y)$ 
defined as
\bea\label{Ddef}
\langle V(p',\epsilon) |\bar q iD_\mu b |\bar B(v)\rangle =
{\cal D}(y) i\varepsilon_{\mu\nu\lambda\sigma} \epsilon_\nu^* p_\lambda
p'_\sigma\,.
\eea
From (\ref{sumeff}) one finds now the following result for the corrected symmetry
relation  (\ref{I})
\bea\label{delta}
\kappa_1(g_+ - g_-) + 2m_B g(y) &=&  - \left( (g_++g_-) + \frac{\bar\Lambda}{m_B}(g_+-g_-) +
2 {\cal D}(y)\right) + {\cal O}(m_b^{-3/2})\,.
\eea
Using the leading order symmetry relations (\ref{II}), (\ref{III}) on the right-hand side,
this can be put in a form similar to (\ref{I})
\bea\label{Icorr}
\mbox{(V-1$'$)}:\qquad \kappa_1 (g_+ - g_-) + 2m_B g =
-2(m_V y - \bar\Lambda) g(y) - \frac{1}{m_B} f(y) - 2 {\cal D}(y) + {\cal O}(m_b^{-3/2})\,.
\eea

The remaining two relations (\ref{II}) and (\ref{III}) can be derived in a similar way, 
starting with the two currents $J_{5\mu} = \bar q i\sigma_{\mu\nu}v^\nu \gamma_5 b$ and
$J'_{5\mu} = (g_{\mu\nu} - v_\mu v_\nu) \bar q \gamma^\nu \gamma_5 b$.
Both these currents can be matched onto HQET operators as in  (\ref{Jeff})
\bea
J^{(\prime )}_{5\mu} = 
\tilde c_0^{(\prime )}(\mu) \bar q\gamma_\mu\gamma_5 h_v + 
\tilde c_1^{(\prime )}(\mu) \bar qv_\mu \gamma_5 h_v
+ \frac{1}{2m_b}\sum_{i=1}^6 \tilde b_i^{(\prime )}(\mu) {\cal J}_{5\mu}^i\,,
\eea
where the dimension-4 operators appearing on the right-hand side are defined 
analogously to (\ref{Ji}) with the substitution $h_v \to \gamma_5 h_v$.
The constraints $v^\mu J_{5\mu} = v^\mu J'_{5\mu} = 0$ give relations
among the Wilson coefficients analogous to (\ref{Jconst})
\bea
\tilde c_0^{(\prime )}(\mu) = \tilde c_1^{(\prime )}(\mu)\,,\qquad
\tilde b_1^{(\prime )}(\mu) + \tilde b_2^{(\prime )}(\mu) = 0\,,\quad 
\tilde b_3^{(\prime )}(\mu) + \tilde b_4^{(\prime )}(\mu) + \tilde
b_6^{(\prime )}(\mu) = 0
\eea

The matching conditions for the current $J_{5\mu} = 
\bar q i\sigma_{\mu\nu}v^\nu \gamma_5 b$ are 
\bea\label{c5match}
& &\tilde c_0(m_b) = \tilde c_1(m_b) = 1 - \frac{\alpha_s C_F}{\pi} - (16.09 
- 1.13 n_\ell + 0.13)\left(\frac{\alpha_s}{\pi}\right)^2\\
& &\tilde b_1(m_b) = 1\,, \tilde b_2(m_b) = -1\,,
\quad \tilde b_{3-6}(m_b) = 0\,,\nonumber
\eea
and for $J'_{5\mu} = (g_{\mu\nu} - v_\mu v_\nu) \bar q \gamma^\nu \gamma_5 b$
\bea\label{c5'match}
& &\tilde c'_0(m_b) = \tilde c'_1(m_b) = 1 - (10.88 - 0.77 n_\ell + 0.11)
\left(\frac{\alpha_s}{\pi}\right)^2\\
& &\tilde b'_1(m_b) = \tilde b'_2(m_b) = -1\,,\quad  
\tilde b'_{3-6}(m_b) = 0\,.\nonumber
\eea
The values of $\tilde c_{0,1}^{(\prime )}(m_b)$ quoted above correspond
to the 't Hooft-Veltman scheme used in \cite{BrGr}.

The two symmetry relations (\ref{II}) and (\ref{III}) are obtained
by taking again a linear combination of the matrix elements of $J_{5\mu}$
and $J'_{5\mu}$, with $\kappa_5$ to be determined below
\bea\label{J5QCD}
& &\langle\rho(p',\varepsilon)|J_{5\mu} - \kappa_5 J'_{5\mu}|\bar B(p)\rangle
= \varepsilon^*_\mu \left( g_+ (m_B+v\cdot p') + g_-(m_B-v\cdot p') -
\kappa_5 f\right)\\
& &+ (v\cdot \varepsilon^*)(p_\mu + p'_\mu) 
\left( -g_+ - m_B h(m_B-v\cdot p') + \frac{1}{2m_B}\kappa_5\left[
f - m_B (m_B-v\cdot p') (a_+-a_-)\right]\right)\nonumber\\
& &+ (v\cdot \varepsilon^*)(p_\mu - p'_\mu) 
\left( -g_- + m_B h(m_B+v\cdot p') + \frac{1}{2m_B}\kappa_5\left[
f + m_B (m_B+v\cdot p') (a_+-a_-)\right]\right)\nonumber\,.
\eea
This relation involves a subtle point which deserves special attention.
Usually, the matrix element 
$\langle \rho|J_{5\mu}|\bar B\rangle$ is computed in terms of the form 
factors of the current $\bar q\sigma_{\mu\nu} b$ with the help of the 
relation $\sigma_{\mu\nu}\gamma_5 = \frac{i}{2}\varepsilon_{\mu\nu\alpha\beta} 
\sigma^{\alpha\beta}$ (corresponding to the convention 
$\varepsilon^{0123} = 1$). 
Beyond tree-level, this relation is only true in the 't Hooft-Veltman
scheme for $\gamma_5$. In any other scheme, there are 
calculable ${\cal O}(\alpha_s)$ corrections to the relation between the 
matrix elements $\langle \bar q\sigma_{\mu\nu} b\rangle$ and 
$\langle \bar q\sigma_{\mu\nu}\gamma_5 b\rangle$, which cancel
the scheme dependence in the individual Wilson coefficients 
$\tilde c^{(\prime )}_{0,1}(m_b)$. For this reason, the use of the 
't Hooft-Veltman scheme will be understood throughout in the following.

The combination of matrix elements (\ref{J5QCD}) is given in the effective 
theory by 
\bea
& &\langle\rho(p',\varepsilon)|J_{5\mu} - \kappa_5 J'_{5\mu}|\bar B\rangle
= (\tilde c_0 - \kappa_5 \tilde c_0')
\left( \langle \bar q\gamma_\mu \gamma_5 h_v\rangle +
\langle \bar q v_\mu \gamma_5 h_v\rangle\right)\\
&+&\frac{1}{2m_b} \left\{
\sum_{\rm a = kin,mag} c_{\rm a}(\tilde c_0 - \kappa_5 \tilde c'_0)\langle T
\{\bar q(\gamma_\mu + v_\mu) \gamma_5 h_v\,, i{\cal L}_{\rm a}\}\rangle +
\sum_{i=1}^6 (\tilde b_i - \kappa_5 \tilde b'_i) 
\langle J_{5\mu}^i\rangle\right\}\nonumber
\eea
Using the tree level values of the Wilson coefficients, the contribution of
the dimension-3 operators vanishes with $\kappa_5=1$, which gives the leading
order symmetry relations (\ref{II}) and (\ref{III}). The same cancellation will
hold true to all orders in $\alpha_s(m_Q)$ provided that $\kappa_5$ is chosen as
$\kappa_5 = \tilde c_0(m_b)/\tilde c'_0(m_b) = 1 -\alpha_s(m_b) C_F/\pi 
+ {\cal O}(\alpha_s^2)$.

The remaining matrix elements of dimension-4 operators can be expressed
in terms of leading order form factors (\ref{VA}), (\ref{T}) and 3 additional
form factors ${\cal D}_1, {\cal D}_\pm$ defined as
\bea
\langle V(p',\varepsilon) |\bar qiD_\mu \gamma_5 h_v|\bar B(v)\rangle =
{\cal D}_1 \varepsilon^*_\mu + {\cal D}_+ (\varepsilon^*\cdot p)(p_\mu +
p'_\mu) + {\cal D}_- (\varepsilon^*\cdot p)(p_\mu -p'_\mu)\,.
\eea
The equation of motion for the heavy quark field $iv\cdot D h_v= 0$ implies a relation
among these form factors, such that only two of them are independent. 

Using standard HQET methods, one finds the following two generalized symmetry
relations, including subleading corrections
\bea\label{IIprel}
& &(g_+ + g_-) m_B + v\cdot p'(g_+ - g_-) - \kappa_5 f = \\
& &\qquad(g_+ + g_-) (\bar\Lambda - v\cdot p') + \frac{1}{m_B}
(g_+ - g_-) (\bar\Lambda v\cdot p' - m_V^2) + \frac{2}{m_B} {\cal D}_1 + 
{\cal O}(m_b^{-3/2})\nonumber
\eea
and
\bea\label{IIIprel}
& &g_+ - g_- + 2m_B^2 h + \kappa_5 m_B (a_+-a_-) =\\
& &\qquad
\frac{\bar\Lambda}{m_B} (g_+ - g_-) - 2m_B h(\bar\Lambda - v\cdot p')
-2({\cal D}_+ - {\cal D}_-) + {\cal O}(m_b^{-3/2})\nonumber\,.
\eea
The first relation can be put in a form similar to (\ref{II}) by inserting 
the lowest order result (\ref{Icorr}) for $g_+ - g_-$. One obtains in this 
way the improved version of the symmetry relation (\ref{II}), including 
subleading $\Lambda/m_b$ and hard gluon corrections
\bea\label{IIcorr}
& &\mbox{(V-2$'$)}:\quad
g_+ + g_- - 2m_V y g - \kappa_5 \frac{1}{m_B}f =\\
& &\qquad\quad -2 \frac{m_V}{m_B}(\bar\Lambda y - m_V) g(y) + \frac{\bar\Lambda}{m_B^2}
f + 
\frac{2}{m_B^2}(
m_V m_B y{\cal D}(y) + {\cal D}_1(y)) + {\cal O}(m_b^{-5/2})\nonumber
\eea
The analog of the symmetry relation (\ref{III}) is obtained in a similar 
fashion from (\ref{IIIprel}) and reads
\bea\label{IIIcorr}
& &\mbox{(V-3$'$)}:\quad
\kappa_5 (a_+ - a_-) + 2m_B h - 2g = 2 (\frac{m_V}{m_B} y - 2\frac{\bar\Lambda}{m_B}) g(y)
+ \frac{1}{m_B^2}f(y)
\\
& &\qquad\qquad  - 2h(\bar\Lambda - m_V y) +
\frac{2}{m_B}({\cal D}(y) - {\cal D}_+(y) + {\cal D}_-(y)) + {\cal O}(m_b^{-5/2})\nonumber
\eea

In conclusion, the leading $\Lambda/m_b$ corrections to the symmetry 
relation (\ref{IV}) for $\bar B\to P$ decays are expressed in terms of
one additional form factor $\delta_+ - \delta_-$, while the symmetry
relations for $\bar B\to V$ decays (\ref{I})-(\ref{III}) require
three additional form factors 
${\cal D}(y)$, ${\cal D}_1(y)$ and ${\cal D}_+(y)-{\cal D}_-(y)$.
All these four quantities are matrix elements of the two local operators
in the effective theory $\bar q iD_\mu (\gamma_5 ) h_v$. 
The relative simplicity of this
result can be appreciated by noting that 22 new matrix elements, both local 
and nonlocal, are required for a complete $1/m_Q$ expansion of the 
individual form factors in $\bar B\to V$ decays \cite{Hu}.
We note also that the local nature of these corrections should make 
their computation on the lattice feasible. 
In the following, we will estimate them using various approximation
schemes and models.\\

The form factors $\delta_\pm(y)$ (\ref{ddef}) relevant for the $\bar B\to P$  
decays can be computed in the soft pion limit if $P$ is one of the
members of the Goldstone bosons octet. We use for this purpose the
chiral perturbation theory for heavy hadrons developed in \cite{hhcpt}.
To lowest order in the chiral expansion, there is a unique operator 
which is a realization of
$\bar q\Gamma iD_\mu h_v$. Such an operator must satisfy the following
conditions: a) transforms as $(\bar 3_L, 1_R) + (1_L, \bar 3_R)$
under $SU_L(3) \times SU_R(3)$, b) transforms in the same way under heavy
quark spin rotations as $\bar q\Gamma h_v$ and c) vanishes upon contraction 
with $v^\mu$. The only operator with these properties is
\bea\label{Omu}
\bar q\Gamma iD_\mu h_v \to \frac{i}{2}\alpha \mbox{Tr }[(\gamma_\mu + v_\mu)
\Gamma H_b (\xi^\dagger_{ba} + \xi_{ba})]
+\frac{i}{2}\alpha \mbox{Tr }[(\gamma_\mu + v_\mu)\gamma_5
\Gamma H_b (\xi^\dagger_{ba} - \xi_{ba})]\,,
\eea
where the Goldstone bosons are described by the field $\xi = 
\exp(\frac{i}{f_\pi} M)$,
and the heavy mesons $\bar B_a^{(*)}$ are contained in the superfield $H_a$ 
defined as in \cite{MaWi}.
The low-energy constant $\alpha$ can be determined by taking the 
$\bar B$-to-vacuum
matrix element of this operator with $\Gamma = \gamma^\mu \gamma_5$, which gives
$\alpha = -\frac16 f_B m_B \bar \Lambda$. 

The form factors $\delta_\pm(y)$ introduced in (\ref{ddef}) can be computed now
to the leading order in the chiral expansion. There are two contributions 
involving the operator (\ref{Omu}), the direct and the pole graphs. Their
computation gives for the $\bar B_d\to \pi^+$ form factors (this is multiplied
with $1/\sqrt2$ for the $B^-\to \pi^0$ case)
\bea
\delta_+ - \delta_- = 
\frac{g\bar\Lambda f_B}{3f_\pi}\cdot\frac{m_B}{v\cdot p' + \Delta} =
\frac{(0.5\pm 0.1)\mbox{ GeV}^2}{v\cdot p' + \Delta}
\eea
with $g$ the $BB^*\pi$ coupling and $\Delta = m_{B^*}-m_B$ the hyperfine 
splitting in the $B$ meson system. This matrix element can be also extracted 
from the current algebra calculation of \cite{BLNN}.
We used in the numerical estimate the recent CLEO measurement \cite{CLEO}
$g = 0.59\pm 0.12$, $\bar\Lambda = 350$ MeV and $f_B = 180$ MeV.

For the $\bar B\to V$ case, no predictions can be obtained from chiral symmetry.
Some information can be obtained however from the constituent quark model.  We will 
show in the following that the form factor ${\cal D}(y)$ defined in (\ref{Ddef}) 
vanishes exactly in the quark model, which suggests that its real value might be 
very small. 
To see this, note that in the quark model the vector meson state is in general
a mixture of $S$-wave and a $D$-wave states ($^3S_1$ and $^3D_1$ in spectroscopic
notation $^{2S+1}L_J$), whereas the $B$ meson state contains only a $S$-wave component
$^1S_0$
\bea
& &|V(\vec 0, m_z = +1)\rangle = c_S \phi_S(|\vec k\,|) |\uparrow\uparrow\rangle\\
& &\qquad +
c_D \phi_D(|\vec k\,|)\left( \frac{1}{\sqrt3} Y_{2,+2}(\hat k) |\downarrow
\downarrow\rangle + \frac{1}{\sqrt6} Y_{2,+1}(\hat k) \cdot \frac{1}{\sqrt2}
|\uparrow\downarrow + \downarrow\uparrow\rangle -
\frac{1}{\sqrt2} Y_{2,0}(\hat k)|\uparrow\uparrow\rangle\right)\nonumber\\
& &|\bar B(v)\rangle = \phi_B(|\vec k\,|)\frac{1}{\sqrt2}
|\uparrow\downarrow - \downarrow\uparrow\rangle
\eea
${\cal D}(y)$ is given by the product of an orbital overlap integral and a spin
matrix element. The latter is only nonvanishing for the $D$-wave component of the
vector meson. However the total result vanishes because of the orthogonality of
the spin wavefunctions
\bea
{\cal D}(y) &\propto & \langle V(\vec q, m_z=+1)|(\vec e_{+1}\times \vec q\,)\cdot
\vec k|\bar B(v)\rangle\\
& &  \propto
\int d^3\vec k 
\phi_{D}(|\vec k+\vec q\,|) Y^*_{2,+1}(\hat k) (\vec e_{+1}\cdot \vec k) 
\phi_B(|\vec k\,|)\times \langle 
\uparrow\downarrow + \downarrow\uparrow | \uparrow\downarrow - \downarrow\uparrow\rangle
= 0\,.\nonumber
\eea

At large recoil $E_V \gg \Lambda_{QCD}$, the form factor ${\cal D}(y)$ can
be also computed in a light-cone expansion in $1/E_V$. This kinematical 
region is at best appropriate to $\bar B\to V$ decays at low values of $q^2$,
which is strictly speaking outside the validity domain of (\ref{Icorr}).
Still, such a calculation is instructive as an order of magnitude estimate.
In this approach, the matrix element
(\ref{Ddef}) is given by a diagram with a hard gluon exchanged between the
current and the spectator quark. We will consider in the following the
$\bar B\to \rho$ case, for which one obtains at lowest order in $1/E_\rho$
\bea\label{pQCD}
{\cal D}(E_\rho) = -\frac{\pi \alpha_s N_c C_F}{E_\rho}
f_B f_\rho^\perp \int_0^\infty dk_+ \frac{\phi_B(k_+)}{k_+}
\int_0^1 \mbox{d} u\frac{\phi_\perp(u)}{1-u}\,,
\eea
where $\phi_B(k_+)$ is the leading twist $B$ meson light-cone wave function,
normalized according to $\int_0^\infty \mbox{d}k_+ \phi_B(k_+) = 1$.
For the $\rho$ meson we kept only the twist-2 chiral-odd wavefunction
$\phi_\perp(u)$ \cite{BaBr} appropriate for a transversely polarized meson.
Since we consider all form factors to be renormalized at $\mu = m_b$,
we will be using this scale \footnote{
A more natural scale in this problem is the virtuality of the exchanged
gluon $\mu^2\simeq 
2E_\rho \Lambda_{QCD} \to m_B\Lambda_{QCD} \simeq 2.5$ GeV$^2$ at $q^2=0$.
Therefore one expects sizable radiative corrections
proportional to $\alpha_s \log(\Lambda_{QCD}/m_b)$, which have been 
neglected in our crude estimate.} 
in the numerical evaluation of (\ref{pQCD}).
Using $f_B = 180$ MeV, $f_\rho^\perp(m_b) = 144$ MeV \cite{BaBr},
together with the asymptotic expression $\phi_\perp (u) = 6u(1-u)$ and
the inverse moment of the $B$ wave function
$\langle (k_+)^{-1}\rangle = 3$ GeV$^{-1}$, we find ${\cal D}(E_\rho) \simeq
- (0.9\mbox{ GeV})/E_\rho$.\\

We turn now to a brief discussion of the implications of these relations. 
The most important result following from (\ref{I}), (\ref{II}) was the 
possibility of extracting the radiative form factor $g_+(y)$ from 
semileptonic decays data. Such analyses have been presented in 
\cite{GMM,Sant,LiWi}. This relation can be now extended to next-to-leading 
order in $\Lambda/m_b$ by combining (\ref{Icorr}) with (\ref{II})
\bea\label{rel}
g_+(y) = -(m_B - \bar\Lambda) g(y) - {\cal D}(y) + {\cal O}(m_b^{-3/2})
\eea
Note that including the subleading term in (\ref{Icorr}) is essential, since it 
is formally of the same order in $\Lambda/m_b$ as the other terms kept.

Numerically, the subleading terms in (\ref{Icorr}) can be significant. For 
illustration, consider the $D\to K^*$ form factors at $y=1$, for which numerical
results are available from the E791 Collaboration $g(1) = -(0.49
\pm 0.04)$ GeV$^{-1}$, $f(1) = (1.9\pm 0.1)$ GeV \cite{E791}. 
From (\ref{rel}) one finds $g^{D\to K^*}_+(1)=(0.74\pm 0.06)$, where we 
used $\bar\Lambda =350$ MeV and ${\cal D}(1)=0$. The subleading correction
in (\ref{Icorr}) contributed about 25\% to this result.
On the other hand, the perturbative effect of hard gluons enters only at 
two-loop order (through $\kappa_1$, given in Eq.~(\ref{kappa1})), 
and therefore can be safely neglected.


Assuming that the vector form factor $g^{D\to V}(y)$ has been measured, the 
relation (\ref{rel}) allows the extraction of $g^{D\to V}_+(y)$ with accuracy
of order $\Lambda^2/m_c^2$. 
Neglecting experimental errors, the theoretical uncertainty in such a
determination of $g_+^{D\to V}(y)$ could be made as small as 
10\%. A precise knowledge of this form factor is relevant for a good 
control of the short-distance matrix element in rare $D$ decays \cite{GGHP}.

The corrected symmetry relations we derived in this Letter can be
expected to help
reduce the theoretical uncertainties in constrained extractions of form 
factors from experimental data, such as those presented in 
\cite{GMM,Sant,LiWi,BuHi}, and in
determinations of CKM parameters with the help of SU(3) and heavy quark
symmetry \cite{Grin,LSW,GrPi}. We will present details of such an investigation
in a separate publication.\\

This work has been supported by the DOE under Grant 
No. DOE-FG03-97ER40546.


\begin{thebibliography}{99}

\bibitem{MaWi} A. Manohar and M. B. Wise, {\em Heavy Quark Physics},
{\em Cambridge Monographs on Particle Physics, Nuclear Physics and Cosmology}, 
Cambridge University Press, Cambridge, U.K., 2000, vol.10.

\bibitem{IW} N. Isgur and M. B. Wise, Phys. Rev. D{\bf 42}, 2388
(1990).

\bibitem{BuDo} G. Burdman and J. F. Donoghue, Phys. Lett. B{\bf 270},
   55 (1991).

\bibitem{LEET} M. Dugan and B. Grinstein, Phys. Lett. B{\bf 255}, 583 (1991);
J. Charles {\em et al}, Phys. Rev. D{\bf 60}, 014001 (1999).

\bibitem{BeFe} M. Beneke and T. Feldmann, Nucl. Phys. B{\bf 592},
   3 (2001).

\bibitem{BFPS} C. W. Bauer, S. Fleming, D. Pirjol and I. W. Stewart,
Phys. Rev. D{\bf 63}, 114020 (2001).

\bibitem{BLNN} G. Burdman, Z. Ligeti, M. Neubert and Y. Nir,
Phys. Rev. D{\bf 49}, 2331 (1994).

\bibitem{Hu} C.-S. Huang, C. Liu and C.-T. Yan, Phys. Rev. D{\bf 62}, 054019
(2000).

\bibitem{LuMa} M. Luke and A. Manohar, Phys. Lett. B{\bf 286}, 348 (1992).

\bibitem{GMM} P. A. Griffin, M. Masip and M. McGuigan, 
   Phys. Rev. D{\bf 50}, 5751 (1994).

\bibitem{Sant} P. Santorelli, Z. Phys. C{\bf 61}, 449 (1994).

\bibitem{LiWi} Z. Ligeti and M. B. Wise, Phys. Rev. D{\bf 60}, 117506
(1999).

\bibitem{BrGr} D. Broadhurst and A. Grozin, Phys. Rev. D{\bf 52}, 4082
(1995); A. Grozin,  Phys. Lett. B{\bf 445}, 165 (1998).

\bibitem{hhcpt} M. B. Wise, Phys. Rev. D{\bf 45}, 2188 (1992);
G. Burdman and J. F. Donoghue, Phys. Lett. B{\bf 280}, 287 (1992);
T. M. Yan et al., Phys. Rev. D{\bf 46}, 1148 (1992) [E: Phys. Rev.
D{\bf 55}, 5851 (1997)].

\bibitem{CLEO} A. Anastassov {\em et al.} (CLEO Collaboration),
Phys. Rev. D{\bf 65} 032003 (2002).


\bibitem{BaBr} P. Ball and V. Braun, Phys. Rev. D{\bf 54}, 2182 (1996).

\bibitem{E791} E. M. Aitala {\em et al.} (E791 Collaboration),
Phys. Rev. Lett. {\bf 80} (1998) 1393.

\bibitem{BuHi} G. Burdman and G. Hiller, Phys. Rev. D{\bf 63}, 113008 (2001).

\bibitem{GGHP} G. Burdman, E. Golowich, J. Hewett and S. Pakvasa,
SLAC-PUB-9057 report, hep-ph/0112235.

\bibitem{Grin} B. Grinstein, Phys. Rev. Lett. {\bf 71}, 3067 (1993).

\bibitem{LSW} Z. Ligeti, I. W. Stewart and M. B. Wise, 
Phys. Lett. B{\bf 420}, 359 (1998).

\bibitem{GrPi} B. Grinstein and D. Pirjol, 
Phys. Rev. D{\bf 62}, 093002 (2000).

\end{thebibliography}
\end{document}